\documentclass[a4paper,twoside]{article}

\usepackage{epsfig}
\usepackage{subfigure}
\usepackage{calc}
\usepackage{amssymb}
\usepackage{amstext} 
\usepackage{amsthm} 
\usepackage{xspace}
\usepackage{focus} 
\usepackage[usenames,dvipsnames]{xcolor} 
\usepackage{textcomp}
\usepackage{url}
\usepackage{array,multirow,graphicx}

\usepackage{SCITEPRESS}      

\begin{document}

\newcommand{\focust}{\textsc{Focus}$^{ST}$} 
\newcommand{\fr}{Flex\-Ray\xspace } 
\newcommand{\fconsts}{\ (c_1,...,c_n \in Config)}
\newcommand{\fconst}{\ (c \in Config)}
\newcommand{\infdisjS}[1]{\textsf{disj}\ensuremath{^\textsf{inf}_\textsf{S}(#1)}}
\newcommand{\infdisj}[1]{\textsf{disj}\ensuremath{^\textsf{inf}(#1)}}
\newcommand{\nmod}[2]{\ensuremath{\mathsf{mod}(#1,#2)}} 

\title{Formal specification of the FlexRay protocol using Focus$^{ST}$}


\author{\authorname{Maria Spichkova}
\affiliation{School of Science, RMIT University, 414-418 Swanston Street, 3001, Melbourne, Australia} 
 \email{maria.spichkova@rmit.edu.au}
}

\abstract{
FlexRay is a communication protocol developed by the FlexRay Consortium.  The core members of the Consortium are  Freescale Semiconductor,  Robert Bosch GmbH, NXP Semiconductors, BMW, Volkswagen, Daimler, and General Motors, and the protocol was respectively oriented towards embedded systems in the automotive domain.
This paper presents a formal specification of the FlexRay protocol using the FocusST framework.   This work extends our previous research of formal specifications of this protocol using Focus formal language. 
}

\onecolumn 
\maketitle 
\normalsize \vfill

\section{\uppercase{Introduction}}

The \fr Consortium developed approx. 20 years ago a time-triggered protocol
for embedded systems in vehicles, cf. \cite{makowitz2006flexray}. The core advantages  of this protocol, 
in comparison for event-driven protocols, 
are deterministic real-time message transmission, fault tolerance, 
integrated functionality for clock synchronisation, and higher bandwidth.
\fr static 
cyclic communication schedules and system-wide synchronous clocks  
allow to apply   distributed control algorithms 
used in drive-by-wire applications.

In our previous work \cite{spichkova2006flexray,kuhnel2006flexray,kuhnel2006upcoming,efts_book}, we introduced the \fr specification using the \Focus specification language \cite{focus} and the corresponding verification using the Isabelle/HOL theorem prover \cite{npw}, cf. also \cite{spichkova2006flexray,spichkova,FocusStreamsCaseStudies-AFP}. That formalization was based on the ``Protocol Specification 2.0''\cite{FlexRayProtocol}.

In this paper we present an extended version of this specification: we apply the \focust\ framework to allow for a better readability as well as highlighting the timing aspects of the specification.  

\emph{Outline:} 
Section \ref{sec:focust} introduces the basic principles of the \focust framework. 
Section \ref{sec:fr} presents the core features of the \fr protocol along with their formal specifications in \focust.
Section \ref{sec:related} discusses the related work. 
Finally, Section \ref{sec:conclusions} summarises the paper.

\section{\uppercase{Focus$^{ST}$}}
\label{sec:focust}

\focust \cite{spichkova2016spatio,spichkova2014modeling} is an extension of the \Focus framework to increase the readability and understandability of the formal specification.  
The \focust\ specification layout   is similar to \Focus, 
but it has a number of new features based on human factor analysis within formal methods, cf.  \cite{spichkova2017human,hffm_spichkova,spichkova2013design}. 

The first step towards elaboration the \focust framework was the optimisation of the \Focus specification layout were discussed in \cite{spichkova2012time}. 
In both frameworks, specifications are based on the notion of \emph{streams}, but the formalisation of this concept is done in different ways:
The input and output streams of a \Focus  component are mappings
 of natural numbers $\Nat$ to single messages and, in the case of timed streams, to $\ntick$ presenting the clock ticks. 
The \focust\  input and output streams of a component are 
 always timed. They are   
formalised as  a mapping from $\Nat$ to lists of messages that are transmitted within the corresponding time intervals. Thus, 
\focust has streams of two kinds ($\nfst{T}$ denotes a list of elements of type $T$):
	\begin{itemize}
	\item
	\emph{Infinite timed streams}  $M^{\underline{\infty}}$ to specify the input and the output streams are formalised by
$
\Nat\ \to \nfst{T}
$;
	\item
	\emph{Finite timed streams}  $M^{\underline{*}}$  to specify timed streams truncated 
	at some point of time are formalised by
$
\nfst{(\nfst{T})}
$.
	\end{itemize}  
    
\section{\uppercase{FlexRay Protocol}}
\label{sec:fr}
 
\begin{figure}[ht!]
   \includegraphics[width=7.5cm]{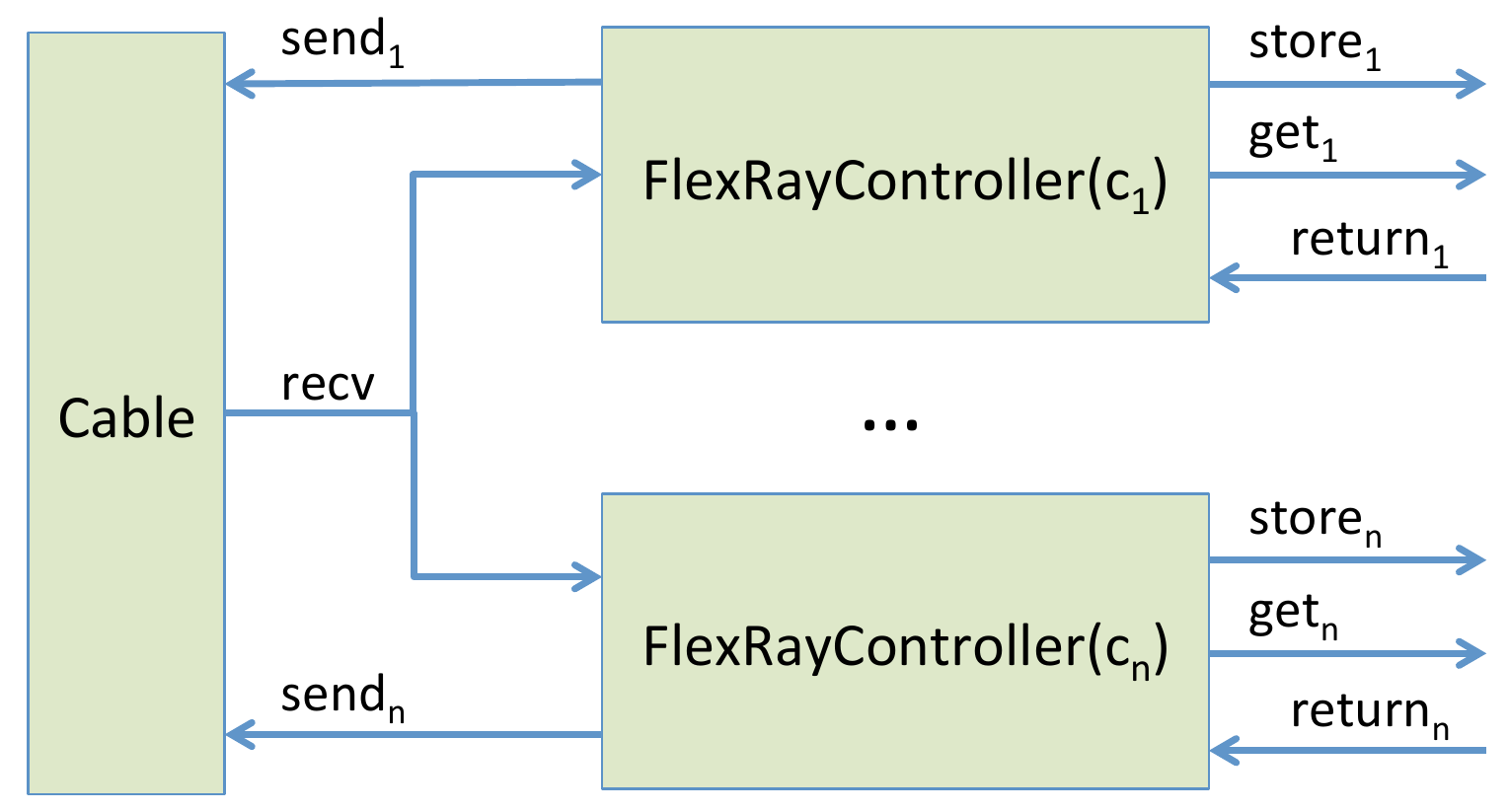}
 \caption{Architecture of the \fr protocol}
  \label{fig:fr}
 \end{figure}

A \fr-based system is built from a number of nodes, connected via a network cable. 
The nodes might have different configurations.
On each node 
\begin{itemize}
\item 
a \fr Controller is running (a network cable connects the FlexRay controllers of all nodes) 
\item 
a number of automotive applications are running. 
\end{itemize}
The \fr message transmission model is based
on \textit{rounds}: each round consist of a constant number of \emph{slots},
time intervals of the same length.
A node can broadcast its messages to other nodes at
statically defined slots. At most one node can
do it during any slot. 
 A high level architecture of \fr\ is presented in Figure \ref{fig:fr}.

\noindent
To specify \fr\ formally, the following data types are defined:
\begin{itemize}
\item
A \fr message consists of a message ID and the actual data that provided by the 
a fault-tolerant communication
layer  FTCom (Fault-Tolerant Communication, cf. \cite{ftcomSpec}) of the 
OSEKtime operationg system \cite{osektime}:
\[
	Message  =  msg~(msgID: \Nat, ftcdata : Data)
\]
\item 
A \fr frame consists of a slot ID and payload (the messages that have to be transmitted), which
is defined as a finite list of type \emph{Message}
\[
	Frame    =   frame (slot : \Nat, data : \nfst{Message})
\]
\item
A \fr bus configuration \emph{Config} onsists of the
scheduling table \emph{schedule} of a node and the length 
of the communication round \emph{cLength}: 
\[
	Config   =   conf~(schedule: \nfst{\Nat}, cLength: \Nat
\]
A scheduling table of a node consists of a number of slots 
in which this node should be sending a frame with the corresponding identifier
(identifier that is equal to the slot).
\end{itemize}
 ~

\subsection{Logical components of the system}
\label{sec:components}

Even having small changes in the syntax might lead to a significant increase of readability and understandability of a formal language. 
For example, numbering the formulas allowing implicit constructs can be very helpful. 
To increase the readability of 
\focust, we use so-called \emph{implicit else-case} constructs:
\begin{itemize}
\item 
if there is no explicit mentioning of the value of an output stream for a particular transaction, then the corresponding time interval of this stream should be empty;
\item
if there is no explicit mentioning of the value of a variable after a particular transaction, then this variable should keep its current value.
\end{itemize}

 \begin{figure}[ht!]
{\footnotesize
\begin{spec}{\spc{FlexRayArch} \fconsts}{td}
\InOutLoc{return_1 ,..., return_n : Frame}{store_1 ,..., store_n : Frame ;	get_1 ,..., get_n : \Nat}{send_1, \dots, send_n, recv : Frame} 
\uasm \\
\cbox{A1} \t1 \forall i \in [1..n]: \msg{1}{return_i}\\
\cbox{A2} \t1  	DisjointSchedules(c_1 ,\dots, c_n)\\
\cbox{A3} \t1  IdenticCycleLength(c_1 ,\dots, c_n) \\
 \zeddashline
\ugar\\
\cbox{G1} \t1 (recv) := Cable(send_1, \dots, send_n)\\
\cbox{G2} \t1 \forall i \in [1..n]:\\
	\t3 (store_i, send_i, get_i) := \\
    \t3 FlexRayController(c_i)(return_i, recv)\\
\end{spec}
} 
 \caption{\fr Architecture specified in \focust}
 \label{fig:frArch}
 \end{figure}

\noindent 
Figure  \ref{fig:frArch} presents 
the architecture of the \fr communication protocol   specified in \focust.  
Its assumption-part consists of three constraints: 
\begin{itemize}
\item[A1:] 
 all bus configurations have disjoint scheduling tables, 
\item[A2:]  all bus configurations have the equal length of the communication round,
\item[A3:]  each controller can receive at most one data frame each time interval from the environment' of the system.
\end{itemize}
The guarantee-part  represents 
architecture of the \fr communication protocol: the system  consists of the component \emph{Cable} 
and $n$ components \emph{FlexRay\_Controller} 
(one controller component for each of  $n$ nodes).

The component \emph{Cable} represents the transmission properties of a
physical network cable:  
every received \fr frame has to be resent to all connected nodes, cf. Figure~\ref{fig:cable}. 
Thus, if one of the controllers send a frame, 
it should be be transmitted to all nodes, i.e., to all other  
controllers in the system.
The specification has only one formula in the assumption part: it expresses 
 that all input streams of this component have to be
disjoint. This assumption is fulfilled due to the properties of the 
\emph{FlexRayController} components and the overall system assumption that 
the scheduling tables of all nodes are disjoint (cf. assumption $A2$ of the \emph{FlexRayArch} specification).  
The guarantee part of this specification has also only one formula:  the predicate \emph{Broadcast} specified in in Section~\ref{sec:aux} below. 

 \begin{figure}[ht!]
{\footnotesize
\begin{spec}{\spc{Cable}}{td}
\InOut{send_1,...,send_n: Frame}{recv: Frame}
\tab{\uasm}\\
\cbox{A1} \t1  \infdisj{send_1,...,send_n}
\zeddashline
\tab{\ugar}	\\
\cbox{G1} \t1  Broadcast(send_1,...,send_n,recv) 
\end{spec}
}
 \caption{Specification of the cable component}
 \label{fig:cable}
 \end{figure}

The specification \emph{FlexRayController} represent the controller component 
for a single node, cf. Figure~\ref{fig:controller}. 
The specification does not have any assumptions on the input streams of this component, which is highlighted by $\ntrue$ in the assumption part.
The guarantee-part  represents 
architecture of this component, as this component is
is a composite one and is built from the components \emph{Scheduler} 
and \emph{BusInterface}. 

 \begin{figure}[ht!] 
{\footnotesize
\begin{spec}{\spc{FlexRayController} (c \in Config)}{td}
\InOutLoc{return, recv : Frame}{store, send : Frame; get : \Nat}{activation : \Nat}
\tab{\uasm} \ntrue
\zeddashline
\tab{\ugar}	\\
\cbox{G1} \t1 (activation) := Scheduler(c)\\
\cbox{G2} \t1  (store, send, get) := \\
	\t3 BusInterface(activation, return, recv)
\end{spec}
}
 \caption{Specification of the controller component}
 \label{fig:controller}
 \end{figure}

\begin{figure}[ht!] 
{\footnotesize
\begin{spec}{\spc{Scheduler} (c \in Config)}{td}
\Out{activation:\Nat}
\tab{univ} s \in \Nat
\zeddashline
\tab{\uasm} \ntrue
\zeddashline
\tab{\ugar}	\\
\cbox{G1} \t1 \forall t \in \Nat:\\
		\t3 s \in schedule(c) \to \ti{activation}{t}=\angles{s}\\
\nwhere s = \nmod{t}{cycleLength(c)}
\end{spec} 
}
 \caption{Specification of the scheduler component}
 \label{fig:scheduler}
 \end{figure} 
 
 \begin{figure}[ht!] 
{\footnotesize
\begin{spec}{\spc{BusInterface}}{td}
\InOut{activation: \Nat; return, recv : Frame}{send, store: Frame; get: \Nat}
\tab{\uasm} \ntrue
\zeddashline
\tab{\ugar}	\\
\cbox{G1} \t1 
\spc{Receive}(recv, store, activation) \\
\cbox{G2} \t1 
\spc{Send}(return, send, get, activation) 
\end{spec}
}
 \caption{Specification of the bus interface component}
 \label{fig:bus}
 \end{figure}

\emph{Scheduler} activates \emph{BusInterface} according to the \fr schedule, cf. Figure~\ref{fig:scheduler}:  
 every time $t$ interval, which is equal (modulo the length 
 of the \fr\ communication cycle) to some frame identifier $i$, the frame with this identifier. The frame identifier 
 corresponds in the scheduler table to the number of the slot in the communication round. 
 
\emph{BusInterface} (cf. Figure~\ref{fig:bus}) specifies
the interaction with other nodes of the \fr\ system,  i.e., 
on what time interval what \fr frame must be send from the node, and how the sended frames should be received. 
The component is specified using by two auxiliary predicates, \emph{Send} and \emph{Receive}, described in Section~\ref{sec:aux}.

\subsection{Auxiliary predicates}
\label{sec:aux}

We define the following auxiliary  predicates to specify the \fr\ protocol: 
\emph{DisjointSchedules}, \emph{IdenticCycleLength}, 
and \emph{FrameTransmission}, \emph{Broadcast}, \emph{Send} and \emph{Receive},  cf. Figure~\ref{fig:auxp}.  

A sheaf of channels of type \emph{Config} 
\begin{itemize}
\item 
fulfils the predicate \emph{DisjointSchedules}, 
if all bus configurations have disjoint scheduling tables;
\item
fulfils the predicate  \emph{IdenticCycleLength}, 
if all bus configurations have the equal length of the communication round.
\end{itemize}

The predicate \emph{FrameTransmission} defines the correct message transmission: 
if the time interval $t$ is equal (modulo the length of the \fr communication round) 
to the element of the scheduler table of the node $k$, then only the node $k$  
is allowed to send data at the $t$th time interval.   
 
The predicate \emph{Broadcast} describes properties of \fr broadcast.  
The predicates \emph{Send} and \emph{Receive} define the \Focus relations on the streams 
to represent respectively data send and data receive by \fr controller.  

\begin{figure}[ht!] 
{\footnotesize
\begin{schema}{\spc{DisjointSchedules}}
\ c_1,...,c_n \in Config \\
\ST
\forall i,j \in [1..n], j \neq i: \\
\t1 \forall x \in \nrng{schedule}(c_i), y \in \nrng{schedule}(c_j): \\
\t2	x \neq y 
\end{schema}
\begin{schema}{\spc{IdenticCycleLength}}
\ c_1,...,c_n \in Config \\
\ST
\forall i,j \in [1..n]: \\
\t1 cycleLength(c_i) = cycleLength(c_j)
\end{schema}
\begin{schema}{\spc{FrameTransmission}}
\ store_1,...,store_n, return_1,...,return_n \in \ntst{Frame} \\
\ get_1,...,get_n \in \ntst{\Nat} \\
\ c_1,...,c_n  \in Config
\ST
\forall t \in \nat, k \in [1..n] : \\
\t1 \nlet~ s = \nmod{t}{cycleLength(c_k)}~~\nin\\
\t2 s \in schedule(c_k)  \to \\
\t3 \ti{get_k}{t}= \angles{s} \ \wedge \\
\t3 \forall j \in [1..n], j \neq k: \ti{store_j}{t} = \ti{return_k}{t} 
\end{schema} 
\begin{schema}{\spc{Broadcast}}
send_1,...,send_n, recv \in \ntst{Frame}
\ST
\forall t \in \nat: \\
\t1  \exists k \in [1...n]: \ti{send_k}{t} \neq \nempty  \to
 \ti{recv}{t} = \ti{send_k}{t}  
\end{schema}
\begin{schema}{\spc{Send}}
return, send \in \ntst{Frame}; get,activation \in \ntst{\Nat}
\ST
\forall t \in \nat: \\
\t1  \ti{activation}{t} \neq \nempty \to \\
\t1  \ti{get}{t} = \ti{activation}{t} \wedge \ti{send}{t} = \ti{return}{t}  
\end{schema}
\begin{schema}{\spc{Receive}}
recv, store \in \ntst{Frame} ; activation \in \ntst{\Nat}
\ST
\forall t \in \nat: \\
\t1 \ti{activation}{t} = \nempty \to \ti{store}{t} = \ti{recv}{t} 
\end{schema}
}
 \caption{Specifications of the auxiliary predicates}
 \label{fig:auxp}
 \end{figure}

\subsection{Specification of requirements}
\label{sec:req}

The  specification \emph{FlexRayReq}  represents requirements on the protocol, cf. Figure~\ref{fig:freq}: 
If 
the scheduling tables are correct in terms of the predicates 
\emph{Disjoint\-Schedules} (all bus configurations have disjoint scheduling tables) 
and \emph{Identic\-Cycle\-Length} (all bus configurations have the equal length of the communication round), and 
also the \fr component receives in every time interval at most one message from each node 
(via channels $return_i$, $1 \le i \le n$), then
\begin{itemize}
	\item 
	the frame transmission must be correct in terms of the predicate 
	\emph{FrameTransmission};
	\item 
	\fr component sends in every time interval at most one message to each node 
 via channels $get_i$ and $store_i$, $1 \le i \le n$).
\end{itemize} 
Please note that the assumption part of this specification is equal to the assumption part of the specification  \emph{FlexRayArch}.

\begin{figure}[ht!] 
{\footnotesize
\begin{spec}{\spc{FlexRayReq} \fconsts}{td}
\InOut{r_1 ,\dots, r_n : Frame}{s_1 ,\dots, s_n : Frame ;	g_1 ,\dots, g_n : \Nat} 
\uasm \\
\cbox{A1} \t1 \forall i \in [1..n]: \msg{1}{r_i}\\
\cbox{A2} \t1  	DisjointSchedules(c_1 ,\dots, c_n)\\
\cbox{A3} \t1  IdenticCycleLength(c_1 ,\dots, c_n) \\
 \zeddashline
\ugar\\
\cbox{G1} \t1
 FrameTransmission(r_1,\dots,r_n,\\
 \t5 s_1,\dots,s_n, g_1,\dots,g_n, c_1,\dots, c_n ) \\
\cbox{G1} \t1 \forall i \in [1..n]:\\
\t4\msg{1}{g_i} \wedge \msg{1}{s_i}
\end{spec}
}
 \caption{Specifications of the \fr\ requirements}
 \label{fig:freq}
 \end{figure}

To demonstrate that the specified \fr\ system fulfils the requirements we need to prove that
the specification \emph{FlexRayArch} is a refinement of the specification \emph{FlexRayReq}.

\section{\uppercase{Related Work}}
\label{sec:related}

\subsection{\focust}

A systematic review of the \Focus related approaches, incl. \focust,  as well as on the case studies they were applied on, was presented in~\cite{spichkova2017focus}. 
Spatio-temporal models for formal analysis and property-based testing were presented in 
\cite{alzahrani2016spatio,alzahrani2017temporal} by Alzahrani et al.
The authors aimed to to apply property-based testing on \focust and TLA models  with temporal properties. 
  Another approach based on \focust, allows analysis of component dependencies    \cite{spichkova2014formalisation}. This was later extended to framework for formal analysis of dependencies among services \cite{spichkova2014towards}.
 
A number of case studies on the modelling of autonomous systems were presented in \cite{spichkova2015towards,spichkova2017autonomous,spichkova2016automotive,spichkova2011decomp}. 
There are also a number of specification ans software development methodologies applying \focust, cf.   
 \cite{botaschanjan2006towards,botaschanjan2005towards,botaschanjan2008correctness,feilkas2009top,feilkas2011refined,spichkova2012cyber,spichkova2013we,holzl2010autofocus,holzl2010safety}. 
An approach introduced by Doby et al. \cite{dobi2015model} utilized \Focus to provide 
an efficient hazard and impact analysis for automotive mechatronics systems.

\subsection{\fr}

Timing analysis of the FlexRay communication protocol were discussed in \cite{pop2008timing}.

Message scheduling for the static and dynamic segments of \fr were analysed in \cite{schmidt2009message1} and \cite{schmidt2009message2} respectively.
There are many approaches on schedule optimization of the static segment, cf.  
\cite{lukasiewycz2009flexray,kang2009static,zeng2011schedule,ding2005ga,schmidt2010optimal,tanasa2010scheduling}.

 A formal verification of the clock synchronization algorithm and of the bus guardian 
of FlexRay was conducted at INRIA~\cite{Zhang}. In our research, we focused on the verification of the communication properties of the protocol. 

Performance analysis of the \fr-based networks was discussed in \cite{hagiescu2007performance,chokshi2010performance}. 
An optimization method for \fr network parameters was proposed in \cite{park2011flexray}. 

A comparison of TTP/C with FlexRay protocol  was presented in \cite{kopetz2001comparison}. 
A comparison of time-triggered Ethernet with FlexRay was introduced in \cite{steinbach2010comparing}. 
CAN, TTCAN, FlexRay and LIN protocols in passenger vehicles were also compared in \cite{talbot2009comparision}.

 An approach for application of time-triggered paradigm (incl. the OSEKtime and \fr aspects) to the 
domain of autonomous systems \cite{spichkova2016automotive}.
An implementation of FlexRay communication controller protocol with application to a robot system was introduced in \cite{xu2008implementation}.

\section{\uppercase{Conclusions}}
\label{sec:conclusions}

This paper presents a formal specification of the \fr protocol using the \focust\ framework.   
This work extends or previous research of formal specifications of this protocol using \Focus formal language and demonstrates the visual improvements as well as the simplifications of the specifications, which initial versions were presented in \cite{spichkova2006flexray,spichkova}.

\bibliographystyle{abbrv} 
{\small

\begin{thebibliography}{10}

\bibitem{alzahrani2016spatio}
N.~Alzahrani, M.~Spichkova, and J.~O. Blech.
\newblock Spatio-temporal models for formal analysis and property-based
  testing.
\newblock In {\em Federation of International Conferences on Software
  Technologies: Applications and Foundations}, pages 196--206. Springer, 2016.

\bibitem{alzahrani2017temporal}
N.~Alzahrani, M.~Spichkova, and J.~O. Blech.
\newblock From temporal models to property-based testing.
\newblock In {\em 11th International Conference on Evaluation of Novel
  Approaches to Software Engineering}, pages 241--246. SCITEPRESS, 2017.

\bibitem{botaschanjan2008correctness}
J.~Botaschanjan, M.~Broy, A.~Gruler, A.~Harhurin, S.~Knapp, L.~Kof, W.~Paul,
  and M.~Spichkova.
\newblock On the correctness of upper layers of automotive systems.
\newblock {\em Formal aspects of computing}, 20(6):637--662, 2008.

\bibitem{botaschanjan2006towards}
J.~Botaschanjan, A.~Gruler, A.~Harhurin, L.~Kof, M.~Spichkova, and
  D.~Trachtenherz.
\newblock Towards modularized verification of distributed time-triggered
  systems.
\newblock In {\em Int. Symposium on Formal Methods}, pages 163--178. Springer,
  2006.

\bibitem{botaschanjan2005towards}
J.~Botaschanjan, L.~Kof, C.~K{\"u}hnel, and M.~Spichkova.
\newblock Towards verified automotive software.
\newblock In {\em ACM SIGSOFT Software Engineering Notes}, volume~30, pages
  1--6. ACM, 2005.

\bibitem{focus}
M.~Broy and K.~St{\o}len.
\newblock {\em Specification and Development of Interactive Systems: Focus on
  Streams, Interfaces, and Refinement}.
\newblock Springer, 2001.

\bibitem{chokshi2010performance}
D.~B. Chokshi and P.~Bhaduri.
\newblock Performance analysis of flexray-based systems using real-time
  calculus, revisited.
\newblock In {\em Proceedings of the 2010 ACM Symposium on Applied Computing},
  pages 351--356. ACM, 2010.

\bibitem{ding2005ga}
S.~Ding, N.~Murakami, H.~Tomiyama, and H.~Takada.
\newblock A ga-based scheduling method for flexray systems.
\newblock In {\em Proceedings of the 5th ACM international conference on
  Embedded software}, pages 110--113. ACM, 2005.

\bibitem{dobi2015model}
S.~Dobi, M.~Gleirscher, M.~Spichkova, and P.~Struss.
\newblock Model-based hazard and impact analysis.
\newblock {\em arXiv preprint arXiv:1512.02759}, 2015.

\bibitem{feilkas2009top}
M.~Feilkas, A.~Fleischmann, F.~H{\"o}lzl, C.~Pfaller, K.~Scheidemann,
  M.~Spichkova, and D.~Trachtenherz.
\newblock A top-down methodology for the development of automotive software.
\newblock {\em Technische Universit{\"a}t M{\"u}nchen, Tech. Rep.}, 902, 2009.

\bibitem{feilkas2011refined}
M.~Feilkas, F.~H{\"o}lzl, C.~Pfaller, S.~Rittmann, B.~Sch{\"a}tz, W.~Schwitzer,
  W.~Sitou, M.~Spichkova, and D.~Trachtenherz.
\newblock {A Refined Top-Down Methodology for the Development of Automotive
  Software Systems -- The KeylessEntry System Case Study}.
\newblock {\em Technische Universit{\"a}t M{\"u}nchen, Tech. Rep.}, 1103, 2011.

\bibitem{FlexRayProtocol}
{FlexRay Consortium}.
\newblock {\em {FlexRay Communication System - Protocol Specification - Version
  2.0}}, 2004.

\bibitem{hagiescu2007performance}
A.~Hagiescu, U.~D. Bordoloi, S.~Chakraborty, P.~Sampath, P.~V.~V. Ganesan, and
  S.~Ramesh.
\newblock Performance analysis of flexray-based ecu networks.
\newblock In {\em Design Automation Conference, 2007. DAC'07. 44th ACM/IEEE},
  pages 284--289. IEEE, 2007.

\bibitem{holzl2010autofocus}
F.~H{\"o}lzl, M.~Spichkova, and D.~Trachtenherz.
\newblock {AutoFocus Tool Chain}.
\newblock {\em Technische Universit{\"a}t M{\"u}nchen, Tech. Rep.},
  (TUM-I1021), 2010.

\bibitem{holzl2010safety}
F.~H{\"o}lzl, M.~Spichkova, and D.~Trachtenherz.
\newblock Safety-critical system development methodology.
\newblock {\em Technische Universit{\"a}t M{\"u}nchen, Tech. Rep.}, (1020),
  2010.

\bibitem{kang2009static}
M.~Kang, K.~Park, and B.~Kim.
\newblock A static message scheduling algorithm for reducing flexray network
  utilization.
\newblock In {\em Industrial Electronics, 2009. ISIE 2009. IEEE International
  Symposium on}, pages 1287--1291. IEEE, 2009.

\bibitem{kopetz2001comparison}
H.~Kopetz.
\newblock A comparison of ttp/c and flexray.
\newblock {\em Institut fur Technische Informatik, Technische Universitat Wien,
  Austria, Research Report}, 10:1--22, 2001.

\bibitem{kuhnel2006flexray}
C.~K{\"u}hnel and M.~Spichkova.
\newblock {FlexRay und FTCom: Formale Spezifikation in FOCUS}.
\newblock {\em Technische Universit{\"a}t M{\"u}nchen, Tech. Rep. I}, 601:2006,
  2006.

\bibitem{kuhnel2006upcoming}
C.~K{\"u}hnel and M.~Spichkova.
\newblock {Upcoming automotive standards for fault-tolerant communication:
  FlexRay and OSEKtime FTCom}.
\newblock In {\em EFTS 2006 International Workshop on Engineering of Fault
  Tolerant Systems. Universite du Luxembourg, CSC: Computer Science and
  Communication}, 2006.

\bibitem{efts_book}
C.~K{\"u}hnel and M.~Spichkova.
\newblock {Fault-Tolerant Communication for Distributed Embedded Systems}.
\newblock In {\em Software Engineering and Fault Tolerance. Software
  Engineering and Knowledge Engineering}, 2007.

\bibitem{lukasiewycz2009flexray}
M.~Lukasiewycz, M.~Gla{\ss}, J.~Teich, and P.~Milbredt.
\newblock Flexray schedule optimization of the static segment.
\newblock In {\em Proceedings of the 7th IEEE/ACM international conference on
  Hardware/software codesign and system synthesis}, pages 363--372. ACM, 2009.

\bibitem{makowitz2006flexray}
R.~Makowitz and C.~Temple.
\newblock Flexray - a communication network for automotive control systems.
\newblock In {\em 2006 IEEE International Workshop on Factory Communication
  Systems}, pages 207--212, 2006.

\bibitem{npw}
T.~Nipkow, L.~C. Paulson, and M.~Wenzel.
\newblock {\em {Isabelle/HOL -- A Proof Assistant for Higher-Order Logic}},
  volume 2283 of {\em LNCS}.
\newblock Springer, 2002.

\bibitem{ftcomSpec}
{OSEK/VDX}.
\newblock {Fault-Tolerant Communication. Specification 1.0}, 2001.

\bibitem{osektime}
{OSEK/VDX}.
\newblock {Time-Triggered Operating System. Specification 1.0}, 2001.

\bibitem{park2011flexray}
I.~Park and M.~Sunwoo.
\newblock Flexray network parameter optimization method for automotive
  applications.
\newblock {\em IEEE Transactions on Industrial Electronics}, 58(4):1449--1459,
  2011.

\bibitem{pop2008timing}
T.~Pop, P.~Pop, P.~Eles, Z.~Peng, and A.~Andrei.
\newblock Timing analysis of the flexray communication protocol.
\newblock {\em Real-time systems}, 39(1-3):205--235, 2008.

\bibitem{schmidt2009message2}
E.~G. Schmidt and K.~Schmidt.
\newblock Message scheduling for the flexray protocol: The dynamic segment.
\newblock {\em IEEE Transactions on Vehicular Technology}, 58(5):2160--2169,
  2009.

\bibitem{schmidt2009message1}
K.~Schmidt and E.~G. Schmidt.
\newblock Message scheduling for the flexray protocol: The static segment.
\newblock {\em IEEE transactions on vehicular technology}, 58(5):2170--2179,
  2009.

\bibitem{schmidt2010optimal}
K.~Schmidt and E.~G. Schmidt.
\newblock Optimal message scheduling for the static segment of flexray.
\newblock In {\em Vehicular Technology Conference Fall (VTC 2010-Fall), 2010
  IEEE 72nd}, pages 1--5. IEEE, 2010.

\bibitem{spichkova2006flexray}
M.~Spichkova.
\newblock {FlexRay: Verification of the FOCUS Specification in Isabelle/HOL. A
  Case Study}.
\newblock {\em Technische Universit{\"a}t M{\"u}nchen, Tech. Rep.}, (602),
  2006.

\bibitem{spichkova}
M.~Spichkova.
\newblock {\em {Specification and Seamless Verification of Embedded Real-Time
  Systems: FOCUS on Isabelle}}.
\newblock PhD thesis, TU M{\"u}nchen, 2007.

\bibitem{spichkova2011decomp}
M.~Spichkova.
\newblock {Architecture: Requirements + Decomposition + Refinement}.
\newblock {\em Softwaretechnik-Trends}, 31:4, 2011.

\bibitem{hffm_spichkova}
M.~Spichkova.
\newblock {Human Factors of Formal Methods}.
\newblock In {\em {IADIS Interfaces and Human Computer Interaction}}. IHCI,
  2012.

\bibitem{spichkova2012time}
M.~Spichkova.
\newblock {Towards Focus on Time}.
\newblock In {\em {12th International Workshop on Automated Verification of
  Critical Systems (AVoCS)}}, 2012.

\bibitem{spichkova2013design}
M.~Spichkova.
\newblock Design of formal languages and interfaces: “formal” does not mean
  “unreadable”.
\newblock In {\em Emerging Research and Trends in Interactivity and the
  Human-Computer Interface}. IGI Global, 2013.

\bibitem{FocusStreamsCaseStudies-AFP}
M.~Spichkova.
\newblock {Stream Processing Components: Isabelle/HOL Formalisation and Case
  Studies}.
\newblock {\em {Archive of Formal Proofs}}, 2013.

\bibitem{spichkova2014formalisation}
M.~Spichkova.
\newblock Formalisation and analysis of component dependencies.
\newblock {\em Archive of Formal Proofs}, 2014.

\bibitem{spichkova2016spatio}
M.~Spichkova.
\newblock {Spatio-temporal features of Focus$^{ST}$}.
\newblock {\em arXiv preprint arXiv:1610.07884}, 2016.

\bibitem{spichkova2017focus}
M.~Spichkova.
\newblock {(Auto) Focus approaches and their applications: A systematic
  review}.
\newblock {\em arXiv preprint arXiv:1711.08123}, 2017.

\bibitem{spichkova2014modeling}
M.~Spichkova, J.~Blech, P.~Herrmann, and H.~Schmidt.
\newblock {Modeling spatial aspects of safety-critical systems with
  Focus$^{ST}$}.
\newblock In {\em MoDeVVa2014}, pages 49--58. CEUR, 2014.

\bibitem{spichkova2012cyber}
M.~Spichkova and A.~Campetelli.
\newblock Towards system development methodologies: From software to
  cyber-physical domain.
\newblock In {\em First International Workshop on Formal Techniques for
  Safety-Critical Systems (FTSCS'12)}, 2012.

\bibitem{spichkova2014towards}
M.~Spichkova and H.~Schmidt.
\newblock Towards logical architecture and formal analysis of dependencies
  between services.
\newblock In {\em The 2014 Asia-Pacific Services Computing Conference}, 2014.

\bibitem{spichkova2015towards}
M.~Spichkova and M.~Simic.
\newblock Towards formal modelling of autonomous systems.
\newblock In {\em Intelligent Interactive Multimedia Systems and Services},
  pages 279--288. Springer, 2015.

\bibitem{spichkova2017autonomous}
M.~Spichkova and M.~Simic.
\newblock Autonomous systems research embedded in teaching.
\newblock In {\em Intelligent Interactive Multimedia Systems and Services},
  pages 268--277. Springer, 2017.

\bibitem{spichkova2017human}
M.~Spichkova and M.~Simic.
\newblock Human-centred analysis of the dependencies within sets of proofs.
\newblock In {\em Knowledge-Based and Intelligent Information \& Engineering
  Systems}, pages 2290--2298. Elsevier, 2017.

\bibitem{spichkova2016automotive}
M.~Spichkova, M.~Simic, and H.~Schmidt.
\newblock From automotive to autonomous: Time-triggered operating systems.
\newblock In {\em Intelligent Interactive Multimedia Systems and Services
  2016}, pages 347--359. Springer, 2016.

\bibitem{spichkova2013we}
M.~Spichkova, X.~Zhu, and D.~Mou.
\newblock Do we really need to write documentation for a system?
\newblock In {\em International Conference on Model-Driven Engineering and
  Software Development (MODELSWARD'13)}, 2013.

\bibitem{steinbach2010comparing}
T.~Steinbach, F.~Korf, and T.~C. Schmidt.
\newblock Comparing time-triggered ethernet with flexray: An evaluation of
  competing approaches to real-time for in-vehicle networks.
\newblock In {\em Factory Communication Systems (WFCS), 2010 8th IEEE
  International Workshop on}, pages 199--202. IEEE, 2010.

\bibitem{talbot2009comparision}
S.~C. Talbot and S.~Ren.
\newblock Comparision of fieldbus systems can, ttcan, flexray and lin in
  passenger vehicles.
\newblock In {\em Distributed Computing Systems Workshops, 2009. ICDCS
  Workshops' 09. 29th IEEE International Conference on}, pages 26--31. IEEE,
  2009.

\bibitem{tanasa2010scheduling}
B.~Tanasa, U.~D. Bordoloi, P.~Eles, and Z.~Peng.
\newblock Scheduling for fault-tolerant communication on the static segment of
  flexray.
\newblock In {\em Real-Time Systems Symposium (RTSS), 2010 IEEE 31st}, pages
  385--394. IEEE, 2010.

\bibitem{xu2008implementation}
Y.-N. Xu, Y.-E. Kim, K.-J. Cho, J.-G. Chung, and M.-S. Lim.
\newblock Implementation of flexray communication controller protocol with
  application to a robot system.
\newblock In {\em Electronics, Circuits and Systems, 2008. ICECS 2008. 15th
  IEEE International Conference on}, pages 994--997. IEEE, 2008.

\bibitem{zeng2011schedule}
H.~Zeng, M.~Di~Natale, A.~Ghosal, and A.~Sangiovanni-Vincentelli.
\newblock Schedule optimization of time-triggered systems communicating over
  the flexray static segment.
\newblock {\em IEEE Transactions on Industrial Informatics}, 7(1):1--17, 2011.

\bibitem{Zhang}
B.~Zhang.
\newblock {On the Formal Verification of the FlexRay Communication Protocol}.
\newblock In {\em Automatic Verification of Critical Systems ({AVoCS})}, pages
  184--189, 2006.

\end{thebibliography}

}

\vfill
 
\end{document}